\documentclass[prl,showpacs,twocolumn,amsmath,amssymb]{revtex4}
\usepackage{graphicx}
\usepackage{color}

\newcommand{\ps}{p^\text{s}}

\def\xa{x(\tau)}

\def\la{\lambda(\tau)}
\def\l{\lambda}

\def\beq{\begin{equation}}
\def\ee{\end{equation}}
\def\bi{\begin {itemize}}
\def\ei{\end{itemize}}

\def\lsim
{\protect \raisebox{-0.75ex}[-1.5ex]{$\;\stackrel{<}{\sim}\;$}}
\def\gsim
{\protect \raisebox{-0.75ex}[-1.5ex]{$\;\stackrel{>}{\sim}\;$}}
\def\lsimeq
{\protect \raisebox{-0.75ex}[-1.5ex]{$\;\stackrel{<}{\simeq}\;$}}
\def\gsimeq
{\protect \raisebox{-0.75ex}[-1.5ex]{$\;\stackrel{>}{\simeq}\;$}}

\def\xa{x(\tau)}

\def\xxx{[x(\tau)|x_0]}
\def\kk{(x,\tau)}

\def\la{\lambda(\tau)}
\def\l{\lambda}

\def\lsa{\tilde\lambda(\tau)}
\def\lt{\lambda_t}

\def\dSt{\dot S_{\rm tot}(\tau)}
\def\dst{\dot s_{\rm tot}(\tau)}
\def\dSm{\dot S_{\rm m}(\tau)}
\def\dsm{\dot s_{\rm m}(\tau)}
\def\m{_{\rm m}}
\def\t{_{\rm tot}}

\def\px{p(x,\tau)}
\def\xu{_{|x(\tau)}}

\def\pnt{p_{n(\tau)}(\tau)}

\def\pjp{p_{n_j^+}}
\def\pjm{p_{n_j^-}}
\def\wpm{w_{n_j^+n_j^-}}
\def\wmp{w_{n_j^-n_j^+}}
\def\njp{{n_j}^+}
\def\njm{{n_j}^-}

\def\tot{_{\rm tot}}
\def\m{_{\rm m}}

\def\beq{\begin{equation}}
\def\ee{\end{equation}}

\def\bi{\begin {itemize}}
\def\ei{\end{itemize}}

\def\F{{\cal F}}

\begin{document}

\title{Entropy production along a stochastic trajectory
and an integral fluctuation theorem}
\author{Udo Seifert}

\affiliation{{II.} Institut f\"ur Theoretische Physik, Universit\"at Stuttgart,
  70550 Stuttgart, Germany}

\begin{abstract}
For stochastic non-equilibrium dynamics like a Langevin equation for a
colloidal particle or a master equation for discrete states, entropy 
production along
a single trajectory is studied. It involves both genuine particle entropy
and entropy production  
in the surrounding medium. The integrated sum of both $\Delta s\t$
is shown to obey a fluctuation theorem $\langle\exp[-\Delta s\t]\rangle =1$
for arbitrary initial conditions and
arbitrary time-dependent driving over a finite time interval.
\end{abstract}

\maketitle

\def\lsim
{\protect \raisebox{-0.75ex}[-1.5ex]{$\;\stackrel{<}{\sim}\;$}}

\def\gsim
{\protect \raisebox{-0.75ex}[-1.5ex]{$\;\stackrel{>}{\sim}\;$}}

\def\lsimeq
{\protect \raisebox{-0.75ex}[-1.5ex]{$\;\stackrel{<}{\simeq}\;$}}

\def\gsimeq
{\protect \raisebox{-0.75ex}[-1.5ex]{$\;\stackrel{>}{\simeq}\;$}}

{\sl Introduction. --} 
Can the notions appearing in
the first and second law of thermodynamics consistently be applied
to mesoscopic non-equilibrium processes like dragging a colloidal
particle  through a viscous fluid \cite{wang02,carb04,mazo99,
zon03}? 
Concerning the
first law, Sekimoto interpreted the terms 
in the standard overdamped Langevin
equation as dynamical increments for applied work, internal energy
and dissipated heat \cite{seki98}. 
For the second law and, in particular, entropy,
a proper formulation and interpretation is more subtle. 
Entropy might be considered as an ensemble property and therefore not 
to be
applicable to a single trajectory. On the other hand, the so-called
fluctuation theorem
\cite{evan93,evan94,gall95,kurc98,lebo99,sear99,croo99,croo00,jarz00,maes02,gasp04,seif05} quite generally 
relates the probability
of entropy generating trajectories to those of entropy annihilating ones 
which requires obviously a definition of entropy on the level of a single
trajectory.
  While for a colloidal particle immersed
in a heat bath  it is
pretty clear what the entropy change of the bath is, it is less obvious
whether or not one should assign an entropy to the particle itself as well.

The purpose of this paper is to show that consequent adaption
of a previously introduced stochastic entropy  \cite{croo99,qian01}
 to the  
trajectory of a colloidal particle
together with the present original
 discussion of its equation of motion yields a consistent interpretation
of entropy production along a single stochastic trajectory. 
Moreover, it leads to a lucid and concise identification of
boundary terms in fluctuation relations. In fact, we
show 
 for arbitrary time-dependent driving that
the total entropy production obeys an
integral fluctuation theorem which is related to but
different from Jarzynski's non-equilibrium
work relation \cite{jarz97}. The present definition of entropy also
implies that the known steady-state fluctuation theorem holds for
finite times rather than in the
long-time limit only as previously in stochastic dynamics
\cite{kurc98,lebo99}.
In a final step, this approach is generalized to arbitrary
driven stochastic dynamics governed by a master equation with time-dependent
rates.

{\sl Entropy along a trajectory. --}
As a paradigm, we consider overdamped motion  $x(\tau)$ of a particle 
with mobility $\mu$ along a one-dimensional
coordinate in the time-interval $0\leq \tau\leq t$
subject to a force
\beq
F(x,\l)=-\partial_xV(x,\l) + f(x,\l) .
\ee
This force can arise from a conservative potential $V(x,\l)$
 and/or be
applied to the particle directly as $ f(x,\l)$. Both sources
may be time-dependent through an external control
parameter $\la$ varied according to some prescribed 
experimental protocol from $\l(0)\equiv \l_0$ to 
$\l(t)\equiv \l_t$. The motion is governed by the Langevin
equation 
\beq
\dot x = \mu F(x,\l) + \zeta,
\ee
with stochastic increments modelled as  Gaussian white noise
with
$
\langle \zeta(\tau)\zeta(\tau')\rangle = 2 D \delta(\tau-\tau')
$
where $D$ is the diffusion constant. In equilibrium, $D$ and $\mu$ 
are related
by the Einstein relation
$
D=T \mu
$ where $T$ is the temperature of the surrounding medium. We assume
this relation to persist even in a non-equilibrium situation.
Throughout 
the paper we set
Boltzmann's constant  to unity such that entropy becomes
dimensionless.

For a definition of entropy along the trajectory, we consider
first the corresponding Fokker-Planck equation  
for the probability $p(x,\tau)$ to find the particle at $x$ at time 
$\tau$
as 
\beq
\partial_\tau p(x,\tau) = - \partial_x j(x,\tau) 
=-\partial_x \left(\mu F(x,\l)-D\partial_xp(x,\tau)\right) .
\label{eq:fp}
\ee
This partial differential equation must be augmented by a
normalized initial
distribution $p(x,0)\equiv p_0(x)$. It will become crucial to distinguish
the dynamical solution $p(x,\tau)$ of this Fokker-Planck equation,
which depends on this given initial condition, from the solution $p^s(x,\l)$ 
for which the rhs of eq. (\ref{eq:fp})
vanishes at any fixed $\l$. The latter corresponds either to 
a steady state for $f\neq0 $ or to  equilibrium for $f=0$,
respectively.

The common definition of a non-equilibrium Gibbs entropy
\beq
S(\tau) \equiv  - \int dx p(x,\tau) \ln p(x,\tau) 
\equiv \langle s(\tau)\rangle 
\label{eq:ens}
\ee suggests to define
 a trajectory-dependent entropy for the particle (or ``system'')
\beq
s(\tau) = -\ln p(x(\tau),\tau)
\label{eq:s}
\ee
where the probability $p(x,\tau)$ obtained by solving the
Fokker-Planck equation is evaluated along the stochastic 
trajectory $x(\tau)$. Obviously, for any given trajectory
$\xa$, the entropy $s(\tau)$  depends on the given initial
data $p_0(x)$ and thus contains information on the
whole ensemble.  For an equilibrium 
Boltzmann distribution at fixed $\l$, this definition assigns 
an entropy 
\beq
s(x)=(V(x,\l)-\F(\l))/T  ,
\ee with the free energy $\F(\l)\equiv -T\int
dx ~\exp[-V(x,\l)/T]$.
 The definition (\ref{eq:s}) has been
used  previously by Crooks for stochastic 
microscopically reversible
dynamics \cite{croo99} and by Qian 
for stochastic dynamics of macromolecules \cite{qian01}.
Neither work, however, discusses the equation of motion
for this stochastic entropy.

{\it Entropy production. --} 
The rate of change of the  entropy (\ref{eq:s}) is given by 
\begin{eqnarray}
\dot s(\tau)&=& -{\partial_\tau p(x,\tau)\over p(x,\tau)}\xu
-{\partial_x \px\over \px}\xu\dot x
\label{eq:dots} \\
&=& -{\partial_\tau \px\over \px}\xu +   
{j(x,\tau)\over D \px}\xu \dot x 
- {\mu F(x,\l)\over D}\xu \dot x .
\nonumber
\end{eqnarray}
The first equality identifies the explicit and the implicit time-dependence.
The second one uses the Fokker-Planck equation. The third term in the
second line can be related to the rate of  heat  dissipation   in the 
medium 
\beq
\dot q(\tau) = F(x,\l)\dot x \equiv T \dsm
\ee
where we identify the exchanged heat with an increase in entropy
of the medium $s\m$ at temperature $T=D/\mu$. 
Then  (\ref{eq:dots}) can be written as a balance equation for 
the trajectory-dependent total
entropy 
production 
\beq
\dst=\dsm+\dot s(\tau) =  - {\partial_\tau \px\over \px}\xu+ 
{j(x,\tau) \over D\px}\xu \dot x ,
\ee
which is our first central result. The first term on the rhs signifies
a change in $\px$ which can be due to a time-dependent $\la$ or, even
at fixed $\l$, due to relaxation from a non-stationary initial state
$p_0(x)\not = p^s(x,\l_0)$.

Upon averaging, the 
total entropy production rate $\dst$ has to become
positive as required by the second law.
This ensemble average proceeds in two steps.
First, we average over all trajectories 
which are at time $\tau$ at a given $x$ leading to
\beq
\langle \dot x|x,\tau\rangle=j(x,\tau)/p(x,\tau) .
\ee
Second, we average over all $x$ 
with $p(x,\tau)$ as
\beq
\dSt \equiv \langle\dst\rangle
= \int dx {j\kk^2\over D p\kk}  \geq 0 ,
\ee
where  equality holds in equilibrium only.
Averaging the increase in entropy of the medium
 along similar lines leads to
\begin{eqnarray}
 \dSm &\equiv&
 \langle  \dsm \rangle =\langle 
F(x,\tau)\dot x\rangle /T\\
&=&  \int dx F(x,\tau)j(x,\tau)/T .
\end{eqnarray}
Hence upon averaging, the increase in entropy of the system itself becomes 
$
\dot S(\tau)\equiv \langle \dot s(\tau)\rangle= \dSt-\dSm $.
On the ensemble level, this balance equation for the averaged quantities
has previously been derived
directly from the ensemble definition (\ref{eq:ens})
\cite{qian01}. The key point of our approach is that 
 we have defined entropy production (or
annihilation) along a single stochastic
trajectory splitting it up into a medium part and
a part of the particle (system). Beyond the conceptual advantage,
this identification facilitates a 
discussion of  fluctuation theorems.

{\sl Fluctuation theorem. --}
Fluctuation theorems derive from the behaviour
 of the weight of a trajectory under ``time-reversal''
which associates with each protocol $\la$ a
reversed one
$
\lsa\equiv \l(t-\tau) 
$
and a reversed trajectory
$
\tilde x(\tau)\equiv x(t-\tau)  .
$
For a given initial value $x_0\equiv x(0)=
\tilde x(t)\equiv \tilde x_t$ and final value $x_t\equiv x(t) =
\tilde x(0)\equiv \tilde
x_0$, the ratio of probabilities of the forward path $p\xxx$
 and of the backward path $\tilde p[\tilde x(\tau)|\tilde x_0]$
can easily be calculated in the path integral representation
of the Langevin equation as \cite{kurc98}
\beq
\ln {p\xxx\over\tilde  p[\tilde x(\tau)|\tilde x_0]} = 
\int_0^tF(x,\tau)\dot x ~d\tau/T =
\Delta s\m .
\ee
If this quantity is combined with
arbitrary normalized distributions for initial and final value
$p_0(x_0)$ and $p_1(\tilde x_0)=p_1(x_t)$, respectively, according to 
\begin{eqnarray}
 R[\xa,\la;p_0,p_1] &\equiv& 
\ln{ p\xxx ~p_0(x_0) \over \tilde p[\tilde x(\tau)|\tilde x_0] 
 ~p_1 (\tilde x_0)} \\&=&  \Delta s\m + \ln  {p_0(x_0) \over
   p_1 (x_t)} ,
\label{eq:R2}
\end{eqnarray}
one easily derives the integral fluctuation relation \cite{maes02}
\begin{eqnarray}
\langle e^{-R} \rangle &\equiv&\sum_{\xa,x_0}  p\xxx ~p_0(x_0) e^{-R}
\nonumber\\
&=&
 \sum_{\tilde x(\tau),\tilde x_0}  \tilde p[\tilde x(\tau)|\tilde x_0] 
 ~p_1 (\tilde x_0) = 1 .
\label{eq:jar}
\end{eqnarray}
Here,  the average is over both initial values drawn from 
the (in principle arbitrary) initial distribution
$p_0(x_0)$ and trajectories $\xa$ determined by
the noise history $\zeta(\tau)$. Since the 
normalized distribution $p_1(x)$ is  arbitrary,
there are, even for fixed $p_0(x)$, an infinity of choices for 
$R$ which obey the
constraint (\ref{eq:jar}) and its  implication
$\langle R\rangle \geq 0$.

At least two choices of $p_1$ have physical meaning in
the present context. First, 
for $p_1(x_t)=p(x,t)$ which is the solution of the 
Fokker-Planck equation for
the given initial distribution $p_0(x_0)$, the definition
(\ref{eq:s}) implies that the last term in
(\ref{eq:R2}) becomes the
entropy change of the particle $\Delta s$ along the trajectory.
Hence  (\ref{eq:jar}) implies 
 the integral  fluctuation theorem 
\beq
\langle e^{-\Delta s\tot} \rangle =1  
\label{eq:R3}
\ee
which is our second main result. 
This integral theorem for $\Delta s\t$ is truely universal
since it holds  for any kind of initial condition
(not only for $p_0(x_0)=p^s(x_0,\lambda_0)$),
any  time-dependence of force and potential, with (for $f=0$) and
without (for $f\not = 0$) detailed balance at fixed $\l$, and  any  length
of trajectory $t$ without the need for waiting for  relaxation.
Crucial for this universality is our identification of the
boundary term in (\ref{eq:R2}) as the change in entropy of the particle.

As a second important choice, for  $f=0$ and 
$p_{0,1} (x) = p^s(x,\l_{0,t})=\exp[-V(x,\l_{0,t})-\F(\l_{0,t})]$,
 one recovers Jarzynski's relation $\langle \exp[-w_d/T]\rangle =1 $
 \cite{jarz97} since
in this case 
\beq
R=\Delta s\m+[V(x_t,\l_t)-V(x_0,\l_0) - \F(\lt)+ \F(\l_0)]/T =w_d/T
\label{eq:R4}
\ee where 
$w_d$ is the part of the work which is irreversibly lost as dissipated
into the medium.
The difference between the two choices  for $p_1(x)$
is subtle but
important. In the first case, the fluctuation theorem
holds for the total entropy change along the
trajectory evaluated at the very end of the protocol. For Jarzynski's
relation, one takes the distribution corresponding to equilibrium
at the final value of $\l$. The difference 
arises from relaxation of the system towards the final
equilibrium state at constant $\lt$ which further increases the 
averaged entropy
of the particle. In fact, 
$p_1(x) = p(x,t)$ is the one choice which leads to
the smallest $\langle R \rangle$ among all possible $p_1(x)$.

For a steady state at constant $\lambda$ and constant force $f\neq 0$
like for motion along a ring with periodic boundary conditions, by choosing
$p_0(x)=p_1(x)=p^s(x)$ in (\ref{eq:R2}), one obtains the stronger
fluctuation relation \cite{croo99,maes02}
\beq
p(-R)/p(R)=e^{-R}  .
\label{eq:ft2}
\ee 
Since with the definition (\ref{eq:s})  the last term
in (\ref{eq:R2}) is again the change in entropy of the system $\Delta s$, 
the quantity $R$ becomes 
 the total entropy
change $\Delta s\t = \Delta s\m+\Delta s$. Hence, one recovers the
fluctuation theorem for the total change in entropy as 
\beq
p(-\Delta s\t)/p(\Delta s\t)=e^{-\Delta s\t}  
\label{eq:pstot}
\ee
even for a finite length $t$ of the trajectories. 
In contrast,
previous derivations
of this genuine fluctuation theorem  within stochastic dynamics 
\cite{kurc98,lebo99}
hold  in the long-time limit only since they 
  implicitly  ignore
what we call $\Delta s$ and consider only $\Delta s\m$.
 Since the former
is bounded for finite potentials, the latter will always win in the long run. 

{\sl Generalizations. -- }
It is obvious that the present discussion holds as well for systems
with more than one degree of freedom obeying  overdamped coupled
Langevin equations. Rather than spelling out the notational
details, we will
now discuss a more general stochastic dynamics on a discrete set $\{n\}$
of states. Again,
we aim at a consistent definition of an entropy along a trajectory without
having available any a priori notion of heat contrary to the colloidal
case above which 
facilitated the identification of entropy production in the medium there.

 Let a transition
between discrete states $m$ and $n$ occur with a rate $w_{mn}(\l)$, which
depends on an externally controlled time-dependent parameter $\l(\tau)$. The
master equation for the time-dependent probability $p_n(\tau)$ then reads
\begin{equation}
  \partial_\tau p_n(\tau) = \sum_{m \not = n}
  [w_{mn}(\l) p_m(\tau) - w_{nm}(\l) p_n(\tau)].
\label{eq:me}
\end{equation}
For a solution, an initial distribution $p_n(0)$ must be specified as well.
As above, the system is
driven externally from $\l_0$ to $\l_t$ according to a protocol $\l(\tau)$.
For any fixed $\l$, there is a steady state $\ps_n(\l)$ which may or may
not obey detailed balance $p_n^s(\l)w_{nm}(\l)=p_m^s(\l)w_{mn}(\l)$.

A stochastic trajectory $n(\tau)$ starts at $n_0$
and jumps at times $\tau_j$ from $n_j^-$ to
$n_j^+$ ending up at $n_t$.
As entropy along this trajectory, we define
\beq
s(\tau)\equiv - \ln \pnt 
\ee
where $\pnt$ is the solution  $p_n(\tau)$ of the master equation (\ref{eq:me})
for a given
initial distribution $p_n(0)$ taken along the specific trajectory $n(\tau)$.
As in the colloidal case, this entropy will depend on the chosen
initial distribution.

The entropy $s(\tau)$ becomes time-dependent due to two sources.
First, even if the system does not jump, $\pnt$ can be time-dependent
 either for time-independent rates due to possible
relaxation from a non-stationary
initial state or, for time-dependent rates, due to the explicit
time-dependence of $\pnt$. Including the jumps, the change of system
entropy reads
\beq
\dot s(\tau) = -{\partial_\tau \pnt\over
\pnt} -\sum_j\delta(\tau-\tau_j)\ln{\pjp\over \pjm} .
\ee Here, and in the remainder, we suppress notationally the
time-dependence of both $p_n(\tau)$ and the rates $w_{nm}(\tau)$
in the jump terms.
We now split up the rhs into a total entropy production and one of
the medium  as follows
\beq
\dst\equiv  -{\partial_\tau \pnt\over
\pnt} -\sum_j\delta(\tau-\tau_j)\ln{\pjp\wpm\over \pjm\wmp} 
\label{eq:stot}
\ee
and
\beq
\dsm\equiv
 -\sum_j\delta(\tau-\tau_j)
\ln{\wpm\over\wmp}  
\label{eq:sm}
\ee
such that the balance
$
\dst=\dot s(\tau) + \dsm
$
holds.

The rational behind the identification (\ref{eq:sm}) for the increase
in entropy of the medium becomes clear after averaging over many
trajectories. For this average, we need the 
probability
for a jump to occur at $\tau=\tau_j$ from $\njm$ to $\njp$
which is
$p_\njm(\tau_j)\wmp(\tau)$. Hence, one gets
\beq
\dSm\equiv \langle\dsm\rangle = 
\sum_{n,k}p_nw_{nk}\ln{w_{nk}\over w_{kn}} ,
\ee
\beq
\dSt\equiv\langle \dst\rangle =
\sum_{n,k}p_nw_{nk}\ln{p_nw_{nk}\over p_kw_{kn}}
\ee
and
\beq
\dot S(\tau)\equiv\langle\dot s(\tau)\rangle = 
\sum_{n,k}p_nw_{nk}\ln{p_n\over p_k}
\ee
such that the global balance $\dSt=\dSm+\dot S(\tau)$ with
$\dSt\geq 0$ is valid.
By averaging our stochastic expressions, we thus
 recover and generalize established results for the
non-equilibrium ensemble entropy balance available so far for the
steady state only \cite{schn76,lebo99,foot2}.

For the fluctuation theorems, the stochastic quantity $R$ 
is derived from the probability $P[n(\tau)|n_0]$ of a trajectory
$n(\tau)$ to occur
under the protocol $\l(\tau)$ 
and the probability $\tilde P[\tilde n(\tau)|\tilde n_0]$
 for the reversed trajectory $\tilde n(\tau)\equiv n(t-\tau)$
to occur under the reversed protocol $\tilde \l(\tau)\equiv
\l(t-\tau)$. With an arbitrary initial distribution $p^0_n$ and an arbitrary
 final distribution $p^1_n$ it becomes
\beq
 R[n(\tau),\la;p^0_n,p^1_n] \equiv 
 \ln{P[n(\tau)|n_0]p^0_{n_0}\over
\tilde P[\tilde n(\tau)|\tilde n_0]p^1_{\tilde n_0}} =
\Delta s\m + \ln{p_{n_0}^0\over p_{n_t}^1} .
\label{eq:R5}
\ee From the infinity of possible fluctuation relations
$\langle\exp[-R]\rangle = 1$, we
choose two important ones. First, for $p_n^0=p_n^s(\l_0$)
and $p_n^1\equiv p_n(t)$, 
the last term in (\ref{eq:R5}) becomes the increase in system entropy
and $R=\Delta s\t$ the total entropy change. Hence, we have again 
the integral theorem (\ref{eq:R3}).
Second, the
 choice of $p_n^1$ that corresponds to Jarzynski's relation in the
colloidal case above 
is implied in the theorem derived in Ref. \cite{seif04}. 
Finally, in a steady state for time-independent rates, by choosing
$p^0_n=p^1_n=p^s_n$, one has the detailed version (\ref{eq:pstot})
 for the total
entropy change for any finite length of the trajectory as exemplified 
for a molecular motor or enzym in Ref. \cite{seif05}.

{\sl Summarizing perspective. -- } We have expressed the entropy production
along a single stochastic trajectory as a sum of an entropy production of the 
system  and of
the  medium  both for a colloidal particle and for
general stochastic dynamics obeying a master equation. The total entropy
production obeys 
an integral fluctuation theorem for arbitrary time-dependent
driving, for  arbitrary initial conditions and any length of trajectories. 
This theorem and
the Jarzynski relation are both shown to be special cases
 of an infinity of
possible fluctuation relations. With the present definition of entropy,
the detailed fluctuation theorem valid in
steady states for the total entropy production  holds even for trajectories of
finite length.

The trajectory dependent entropy of the particle could be measured 
experimentally for a
time-dependent protocol
by first recording over many 
trajectories
the probability distribution $\px$   from which the entropy $s(\tau)$ 
of each trajectory can be inferred.
With
such data, one could also test the new integral fluctuation theorem 
(\ref{eq:R3})
and compare it to Jarzynski's relation for the same protocol. 
It will be interesting to derive, both
experimentally and theoretically, the probability distribution of these
entropy changes and to see, e.g.,  whether they are Gaussian for slow 
driving as is
the dissipated work appearing in Jarzynski's relation 
\cite{spec04}.


\begin{thebibliography}{23}
\bibitem{foot2}{In \cite {lebo99} a ``microscopic entropy flux'' 
$ \protect \mathaccentV {dot}05Fs_{\protect \rm  m}(\tau )= 
 \sum_m w_{n(\tau )m} \protect \qopname  \relax o{ln}
{w_{n(\tau )m}\over w_{m n(\tau )}} $ was proposed which mixes trajectory 
($n(\tau )$) and ensemble ($\sum_m$) aspects.}
\end{thebibliography}
\end{document}